\documentclass[prb,10pt,twocolumn]{revtex4}

\usepackage{epsfig}
\graphicspath{{/}}

\newcommand{\ul}[1]{\underline{#1}}

\begin{document}

\title{Radiative Lifetimes of Excitons and Trions in Monolayers of Metal Dichalcogenide MoS$_{2}$}

\author{Haining Wang, Changjian Zhang, Weimin Chan, Christina Manolatou, Sandip Tiwari, Farhan Rana}
\affiliation{School of Electrical and Computer Engineering, Cornell University, Ithaca, NY 14853}
\email{fr37@cornell.edu}

\begin{abstract}
We present results on the radiative lifetimes of excitons and trions in a monolayer of metal dichalcogenide MoS$_{2}$. The small exciton radius and the large exciton optical oscillator strength result in radiative lifetimes in the 0.18-0.30 ps range for excitons that have small in-plane momenta and couple to radiation. Average lifetimes of thermally distributed excitons depend linearly on the exciton temperature and can be in the few picoseconds range at small temperatures and more than a nanosecond near room temperature. Localized excitons exhibit lifetimes in the same range and the lifetime increases as the localization length decreases. The radiative lifetimes of trions are in the hundreds of picosecond range and increase with the increase in the trion momentum. Average lifetimes of thermally distributed trions increase with the trion temperature as the trions acquire thermal energy and larger momenta. We expect our theoretical results to be applicable to most other 2D transition metal dichalcogenides.   
\end{abstract}
                                    
\maketitle

\section{Introduction}
The unique electrical and optical properties of Two dimensional (2D) metal dichalcogenides (TMDs) have made them interesting for a variety of different applications~\cite{fai10,fai12,fai13,xu13,wang10,kis11,kis13}. Particularly distinguishing features of 2D metal dichalcogenides are the large exciton and trion binding energies in these materials. The exciton and trion binding energies in 2D chalcogenides are almost an order of magnitude larger compared to other bulk semiconductors \cite{fai10,fai13,xu13,Changjian14,timothy,Chernikov14}. The large exciton and trion binding energies imply that many body interactions play an important role in determining the optical properties of these materials. For most optoelectronic applications, knowing the radiative lifetimes of elementary excitations is critical. Excitons and trions in metal dichalcogenides have small radii and, therefore, large oscillator strengths and, consequently, one expects their radiative lifetimes to be particularly short.

In recent work~\cite{Changjian14}, we quantitatively determined the oscillator strengths of excitons and trions in monolayer metal dichalcogenide MoS$_{2}$ from the experimentally measured optical absorption spectra. Our work showed that the traditional Wannier-Mott exciton model~\cite{haugbook}, with a few modifications, is able to describe the excitons and trions fairly well in 2D dichalcogenides. The applicability of the Wannier-Mott model for excitons in dichalcogenides was also found recently by Chernikov et~al.~\cite{Chernikov14}. In this paper we report the radiative lifetimes of excitons and trions.  

Our results show that the small exciton radius and the large exciton optical oscillator strength in MoS$_{2}$ monolayers result in radiative lifetimes in the 0.18-0.30 ps range for small in-plane momentum excitons. These lifetimes are almost two orders of magnitude shorter than those of 2D excitons in III-V semiconductor quantum wells~\cite{Andreani91}. Average exciton lifetimes in a thermal ensemble depend linearly on the exciton temperature, as is expected for excitons in 2D~\cite{Andreani91}. Localized excitons exhibit lifetimes that increase with the decrease in the localization length $L_{c}$ as $\sim$$L_{c}^{-2}$ and can exceed a nanosecond for localization lengths smaller than $\sim$1 nm. In the case of trions we find that the radiative lifetimes are in the hundreds of picosecond range and increase with the increase in the trion momentum. Average lifetimes of thermally distributed trions increase with the trion temperature as the trions acquire thermal energy and larger momenta. We also find that in highly doped samples at low temperatures the average trion lifetimes can become very long because the carrier (electron or hole) that is left behind when a trion recombines has difficulty finding an unoccupied state in the band. Our results also show that in many commonly encountered experimental situations the ensemble averaged radiative lifetimes of thermally distributed excitons and trions can be comparable. In the main text of the paper we have used Fermi's golden rule  to obtain radiative rates. In the Appendix, we show that a many-body Green's function approach gives the same results and also shows that, contrary to recent suggestions~\cite{Yao14,Qiu15}, exciton dispersion within the light cone is not modified as a result of long-range exchange interactions.     

Although the discussion in this paper focuses on MoS$_{2}$ monolayers, the analysis and the results presented here are expected to be relevant to all 2D metal dichalcogenides (TMDs), and are expected to be useful in designing metal dichalcogenide optoelectronic devices as well as in helping to understand and interpret experimental data~\cite{Shi13,Lagarde14,Korn11}. Recent experimental results~\cite{Huber15,Moody15,Marie15} on the lifetimes of excitons in TMD monolayers have yielded numbers that are in good agreement with the numbers calculated in this paper.

\section{Energy Bands in MoS$_{2}$}
The valence band maxima and conduction band minima in a MoS$_{2}$ monolayer occur at the $K$ and $K'$ points in the Brillouin zone. Most of the weight in the conduction and valence band Bloch states near the $K$ and $K'$ points resides on the d-orbitals of $M$ atoms~\cite{Falko13,yao12,timothy}. The spin up and down valence bands are split near the  $K$ and $K'$ points by 0.1-0.2 eV due to the spin-orbit-coupling~\cite{Falko13,yao12,timothy,Liu14}. In comparison, the spin-orbit-coupling effects in the conduction band are much smaller~\cite{Liu14}. Assuming only d-orbitals for the conduction and valence band states, and including spin-orbit coupling, one obtains the following simple spin-dependent tight-binding Hamiltonian (in matrix form) near the $K$($K'$) points \cite{yao12},
\begin{equation}
\left[
\begin{array}{cc}
\Delta/2 & \hbar v k_{-} \\
\hbar v k_{+} & -\Delta/2 + \lambda \tau \sigma
\end{array} \right]    \label{eq:H1}
\end{equation}
Here, $\Delta$ is related to the material bandgap, $\sigma=\pm1$ stands for the electron spin, $\tau=\pm1$ stands for the $K$ and $K'$ valleys, $2\lambda$ is the splitting of the valence band due to spin-orbit coupling, $k_{\pm}=\tau k_{x}\pm ik_{y}$, and the velocity parameter $v$ is related to the coupling between the orbitals on neighboring $M$ atoms. From density functional theories \cite{Lam12,Falko13}, $v\approx 5-6 \times 10^5 $ m/s. The wavevectors are measured from the $K$($K'$) points. The d-orbital basis used in writing the above Hamiltonian are $| d_{z^{2}} \rangle$ and $(| d_{x^{2}-y^{2}}\rangle + i\tau | d_{xy}\rangle)/\sqrt{2}$~\cite{yao12}. We will use the symbol $s$ for the combined valley ($\tau$) and spin ($\sigma$) degrees of freedom. Defining $\Delta_{s}$ as $\Delta - \lambda \tau \sigma$, the energies and eigenvectors of the conduction and valence bands are~\cite{yao12,Efimkin13},
\begin{equation}
E_{{c \atop v},s}(\vec{k}) = \frac{\lambda \tau \sigma}{2} + \gamma \sqrt{ (\Delta_{s}/2)^{2} + (\hbar v k)^{2}} \label{eq:engydis}
\end{equation}
\begin{equation}
| v_{{c \atop v},\vec{k},s}\rangle  = \left[ \begin{array}{c} \gamma \cos(\theta_{\gamma,\vec{k},s}/2) e^{-i\tau\phi_{\vec{k}}/2} \\ \tau  \sin(\theta_{\gamma,\vec{k},s}/2) e^{i\tau\phi_{\vec{k}}/2} \end{array} \right] \label{eq:wf1}
\end{equation}
Here, $\gamma =1$ (or $-1$) stands for the conduction (or the valence) band, $\phi_{\vec{k}}$ is the phase of the wavevector $\vec{k}$, and,
\begin{equation}
\cos(\theta_{\gamma,\vec{k},s}) = \gamma \frac{\Delta_{s}}{2 \sqrt{ (\Delta_{s}/2)^{2} + (\hbar v k)^{2}}} = \gamma \alpha_{\vec{k},s} \label{eq:cs}
\end{equation}
Near the conduction band minima and valence band maxima, the band energy dispersion is parabolic with effective masses, $m_{e}$ and $m_{h}$, for electrons and holes, respectively. The momentum matrix element between the conduction and valence band states near $K$($K'$) points follows from the above Hamiltonian,
\begin{eqnarray}
  \vec{P}_{vc,s}(\vec{k},\vec{k}) & = & \langle v_{v,\vec{k},s}| \hat{\vec{P}} | v_{c,\vec{k},s}\rangle \nonumber \\
  & = & m_{o}v \, \hat{x} \, \left[ \alpha_{\vec{k},s} \cos(\tau \phi_{\vec{k}}) - i\sin(\tau \phi_{\vec{k}}) \right]  \nonumber \\ 
  & & + i \tau \, m_{o}v \, \hat{y} \, \left[ \cos(\tau \phi_{\vec{k}}) - i \alpha_{\vec{k},s} \sin(\tau \phi_{\vec{k}}) \right] \label{eq:matrix}
\end{eqnarray}
Here, $m_{o}$ is the free electron mass. Near the band extrema, $\alpha_{\vec{k},s} \rightarrow 1$ and $\vec{P}_{vc,s}(\vec{k},\vec{k}) \rightarrow m_{o}v \,(\hat{x} + i\tau \hat{y})e^{-i\tau \phi_{k}}$. The Hamiltonian given above is accurate only near the band edges. Later in this paper, we will need to modify the Hamiltonian and obtain results that are also accurate for large wavevectors.

\section{Exciton States}

\subsection{The Electron and Hole Hamiltonian}
The Hamiltonian describing the electron states in the conduction and valence bands is~\cite{Wang15b}, 
\begin{eqnarray}
H_{o} & = & \sum_{\vec{k},s}E_{c,s}(\vec{k}) c^{\dagger}_{\vec{k},s} c_{\vec{k},s} + \sum_{\vec{k},s}E_{v,s}(\vec{k}) b^{\dagger}_{\vec{k},s} b_{\vec{k},s} + H_{eh}
\end{eqnarray}
Here, $c_{\vec{k},s}$ and $b_{\vec{k},s}$ are the destruction operators for the conduction band and valence band, respectively. The Coulomb interaction $H_{eh}$ between the electrons and holes is,
\begin{equation}
H_{eh} =  \frac{1}{A} \sum_{\vec{k},\vec{k}',\vec{q},s,s'} V(\vec{q})  F_{s,s'}(\vec{k},\vec{k}',\vec{q}) c^{\dagger}_{\vec{k}+\vec{q},s}  b^{\dagger}_{\vec{k}'-\vec{q},s'} b_{\vec{k}',s'} c_{\vec{k},s}
\end{equation}
$V(\vec{q})$ is the 2D Coulomb potential and equals $e^{2}/2\epsilon_{o} \epsilon(\vec{q})q$. The wavevector-dependent dielectric constant $\epsilon(\vec{q})$ for monolayer MoS$_{2}$ is given by Zhang et~al.~\cite{Changjian14} and  Berkelbach et~al.~\cite{timothy}. $F_{s,s'}(\vec{k},\vec{k}',\vec{q})$ is~\cite{Efimkin13},
\begin{equation}
F_{s,s'}(\vec{k},\vec{k}',\vec{q}) = \langle v_{c,\vec{k}+\vec{q},s}| v_{c,\vec{k},s} \rangle \, \langle v_{v,\vec{k}'-\vec{q},s'} | v_{v,\vec{k}',s'} \rangle \label{eq:F1}
\end{equation}
Near the band extrema, where $\hbar v k << \Delta_{s}$, one obtains~\cite{Wang15b,Efimkin13},
\begin{equation}
F_{s,s'}(\vec{k},\vec{k}',\vec{q}) = e^{i(\tau\phi_{\vec{k}+\vec{q}} - \tau\phi_{\vec{k}} + \tau'\phi_{\vec{k}'} - \tau'\phi_{\vec{k}'-\vec{q}})/2} \label{eq:F2}
\end{equation}
Given the large bandgap of MoS$_{2}$, the above approximation for the Coulomb matrix element has been found numerically to be adequate to describe the lowest bound exciton energy state in MoS$_{2}$.   

\subsection{Exciton States with Non-Zero Momentum}
We assume that the ground state $|\psi_{o}\rangle$ of the semiconductor monolayer consists of a completely filled valence band and a conduction band with an electron density $n_{e}$ distributed according to the Fermi-Dirac distribution $f_{c}(\vec{k})$. The ground state thus belongs to a thermal ensemble.  A bright exciton state with in-plane momentum $\vec{Q}$ can be constructed from the ground state as follows~\cite{Wang15b,Efimkin13}, 
\begin{equation}
|\psi_{\vec{Q},s}\rangle_{ex} =   \frac{1}{\sqrt{A}}\sum_{\vec{k}} \frac{\psi_{\vec{Q},s}(\vec{k})}{N_{\vec{Q}}(\vec{k})}  c_{\vec{k} + \lambda_{e} \vec{Q},s}^{\dagger} b_{\vec{k} - \lambda_{h} \vec{Q},s}|\psi_{o} \rangle  \label{eq:exciton}
\end{equation} 
Here, $\psi_{\vec{Q},s}(\vec{k})$ is the exciton wavefunction. $\lambda_{e} + \lambda_{h} = 1$ in order to ensure that the above state has a momentum $\vec{Q}$. In the case of parabolic bands, $\lambda_{e} = m_{e}/m_{ex}$ and $\lambda_{h} = m_{h}/m_{ex}$, where $m_{ex} = m_{e} + m_{h}$ is the exciton mass. These values ensure minimum coupling between the exciton relative and center-of-mass wavefunctions. For non-parabolic bands, these expressions for $\lambda_{e \atop h}$ still hold provided the average values of the inverse electron and hole effective masses with respect to the exciton relative wavefunction are used, as described by Siarkos et~al.~\cite{Siarkos00}. For conduction and valence bands with symmetric energy band dispersions, as in (\ref{eq:engydis}),  $\lambda_{e} = \lambda_{h} = 0.5$. The normalization factor $N_{\vec{Q}}(\vec{k})$ equals $\sqrt{1 - f_{c}(\vec{k}+\lambda_{e}\vec{Q})}$. The exciton state is normalized such that $\left\{ _{ex}\langle \psi_{\vec{Q},s}|\psi_{\vec{Q},s} \rangle_{ex} \right\}_{th} =1$, where the curly brackets represent averaging with respect to the thermal ensemble. This normalization gives, 
\begin{equation}
\int \frac{d^{2}\vec{k}}{(2\pi)^{2}} \psi^{*}_{\vec{Q},s}(\vec{k}) \psi_{\vec{Q},s}(\vec{k}) = 1
\end{equation}
Using the exciton state given in (\ref{eq:exciton}) as a variational state, the exciton wavefunction $\psi_{\vec{Q},s}(\vec{k})$ is found to satisfy the Hermitian eigenvalue equation,
\begin{eqnarray}
& &   \left[\bar{E}_{c,s}(\vec{k}+\lambda_{e}\vec{Q}) - \bar{E}_{v,s}(\vec{k}-\lambda_{h}\vec{Q}) \right]\psi_{\vec{Q},s}(\vec{k})  \nonumber \\
& &  - \frac{\sqrt{1 - f_{c}(\vec{k}+\lambda_{e}\vec{Q})}}{A} \sum_{\vec{q}} V(\vec{q}) F_{s,s}(\vec{k} - \vec{q} + \lambda_{e}\vec{Q},\vec{k}-\lambda_{h}\vec{Q},\vec{q}) \nonumber \\
& &  \times \psi_{\vec{Q},s}(\vec{k}-\vec{q}) \sqrt{1 - f_{c}(\vec{k} - \vec{q} + \lambda_{e}\vec{Q})}  \nonumber \\
  & &  = E_{ex,s}(\vec{Q}) \psi_{\vec{Q},s}(\vec{k}) \nonumber \\
  \label{eq:eigen0}
\end{eqnarray}
The Coulomb potential term is gauge-dependent, a fact that has often been overlooked in the literature. If one writes the exciton wavefunction as,
\begin{equation}
\psi_{\vec{Q},s}(\vec{k}) = \phi_{\vec{Q},s}(\vec{k}) e^{i(\tau\phi_{\vec{k}+ \lambda_{e} \vec{Q}} + \tau\phi_{\vec{k} - \lambda_{h} \vec{Q}})/2} \label{eq:psiphi}
\end{equation}
then, assuming the approximation in (\ref{eq:F2}) for $F$, the exciton wavefunction $\phi_{\vec{Q},s}(\vec{k})$ satisfies the  eigenvalue equation~\cite{Wang15b,Changjian14,timothy},
\begin{eqnarray}
  & & \left[\bar{E}_{c,s}(\vec{k}+\lambda_{e}\vec{Q}) - \bar{E}_{v,s}(\vec{k}-\lambda_{h}\vec{Q}) \right]\phi_{\vec{Q},s}(\vec{k}) \nonumber \\
  & & - \frac{\sqrt{1 - f_{c}(\vec{k}+\lambda_{e}\vec{Q})}}{A} \sum_{\vec{q}} V(\vec{q}) \phi_{\vec{Q},s}(\vec{k}-\vec{q}) \nonumber \\
  & & \times \sqrt{1 - f_{c}(\vec{k} - \vec{q} + \lambda_{e}\vec{Q})} \nonumber \\
& & = E_{ex,s}(\vec{Q}) \phi_{\vec{Q},s}(\vec{k}) \label{eq:eigen}
\end{eqnarray}
with an eigenvalue $E_{ex,s}(\vec{Q})$. {\em Note that all the phase factors cancel out and do not appear in the exciton eigenvalue equation}. The probability of finding an electron and a hole at a distance $\vec{r}$ from each other in the exciton state $| \psi_{\vec{Q},s} \rangle_{ex}$ is $|\phi_{\vec{Q},s}(\vec{r})|^{2}$, where $\phi_{\vec{Q},s}(\vec{r})$ is the Fourier transform of $\phi_{\vec{Q},s}(\vec{k})$, and not of $\psi_{\vec{Q},s}(\vec{k})$, which also includes extra phase factors (see (\ref{eq:psiphi})). Although the lowest energy exciton state (1s) considered in this paper is not much affected by the Coulomb matrix element approximation in (\ref{eq:F2}), the exact expression for the Coulomb matrix element is needed to describe the effects of the Berry's phase on the higher exciton energy levels and the resulting small energy splittings between the exciton states of opposite angular momentum~\cite{Zhou15,Srivastava15}. 

If the exciton momentum $\vec{Q}$ is assumed to be small (only excitons with very small momenta couple to radiation), the $\vec{Q}$ dependence of the Fermi distributions can be ignored in (\ref{eq:eigen}) and the band energy dispersions can be Taylor expanded to express $E_{ex,s}(\vec{Q})$ as,
\begin{equation}
E_{ex,s}(\vec{Q}) = E_{g,s} - E_{exb} + \frac{\hbar^{2} Q^{2}}{2 m_{ex}}
\end{equation}
where, $E_{exb}$ is the exciton binding energy and the bandgap is $E_{g,s} = E_{c,s}(\vec{k}=0) - E_{v,s}(\vec{k}=0)$. The energy $E_{ex,s}(\vec{Q})$ is measured with respect to the energy of the ground state $|\psi_{o}\rangle$. The spin-valley subscript can be dropped from $E_{ex,s}(\vec{Q})$ without causing confusion since the fundamental exciton energies are the same in the two valleys in TMDs.

\subsection{Exciton Exchange Interactions, Self-Energy, and Dispersion}
Recently, it was suggested that exciton dispersion in TMD monolayers is modified as a result of exchange interactions resulting in Dirac cone like features~\cite{Yao14,Qiu15}. The most important exciton exchange interactions are long-range dipole-dipole interactions and require retarded electromagnetic potentials for an accurate description. In Appendix~\ref{app1}, we use a Green's function approach and show that the dispersion relation of excitons is not modified within the light cone as a result of these interactions and, therefore, exchange interactions can be ignored when calculating the exciton radiative lifetimes.

\section{Photon Emission by Excitons} \label{sec:excitons}

\subsection{Hamiltonian for Interaction with Photons} \label{sec:int}
The quantized radiation field is~\cite{haugbook},
\begin{eqnarray}
  \vec{A}(\vec{r}) & = & \sum_{\vec{q},j} \sqrt{\frac{\hbar}{2\epsilon_{o} \omega_{q}}} \left[ \hat{n}_{j}(\vec{q}) a_{j}(\vec{q}) +  \hat{n}^{*}_{j}(-\vec{q}) a^{\dagger}_{j}(-\vec{q}) \right] \frac{e^{i\vec{q}.\vec{r}}}{\sqrt{V}} \nonumber \\
  & = & \sum_{\vec{q},j} \vec{A}_{\vec{q},j} \frac{e^{i\vec{q}.\vec{r}}}{\sqrt{V}}
\end{eqnarray}
Here, $\hat{n}_{j}$ for $j=1,2$ are the field polarization vectors and $a_{j}(\vec{q})$ is the field destruction operator for a mode with wavevector $\vec{q}$ and frequency $\omega_{q}$. The interaction between the electrons and photons is given by the Hamiltonian,
\begin{equation}
H_{int} =  \sum_{\vec{q},j} H^{+}_{j}(\vec{q}) +  H^{-}_{j}(\vec{q}) \nonumber \\
\end{equation}
where,
\begin{eqnarray}
  & &   H^{-}_{j}(\vec{q})  =  \frac{e}{m_{o}\sqrt{V}} \vec{A}_{\vec{q},j} \, \cdotp \, \sum_{\vec{k},s} \vec{P}_{vc,s}(\vec{k}+\vec{q}_{\parallel}, \vec{k})  b^{\dagger}_{\vec{k} + \vec{q}_{\parallel},s}  c_{\vec{k},s} \nonumber \\
  \label{eq:light_H2}
\end{eqnarray} 
and $ H^{+}_{j}(\vec{q}) =  [H^{-}_{j}(\vec{q})]^{\dagger}$. We have expressed the field wavevector $\vec{q}$ in terms of the in-plane component, $\vec{q}_{\parallel}$, and the out-of-plane component, $q_{z}\hat{z}$. The interaction Hamiltonian used above ignores the quadratic vector potential term~\cite{Girlanda95}(see also Appendix~\ref{app1}). This omission is justified as long as one is interested in a resonant linear optical processes.

\subsection{Radiative Lifetime of Excitons}
We assume a suspended MoS$_{2}$ monolayer located in the plane $z=0$. Without losing generality, we restrict ourselves to the valley $\tau=1$ where the top most valence band is occupied by spin-up ($\sigma=1$) electrons, and we will drop the spin-valley index in the discussion that follows. TMDs can have both bright and dark excitons. Dark exciton states do not couple directly to radiation and will not be considered here. We assume an initial state $|\psi_{\vec{Q}}\rangle_{ex}$ consisting of a bright exciton with in-plane momentum $\vec{Q}$,
\begin{equation}
|\psi_{\vec{Q}}\rangle_{ex} =   \frac{1}{\sqrt{A}}\sum_{\vec{k}} \frac{\psi_{\vec{Q}}(\vec{k})}{N_{\vec{Q}}(\vec{k})}  c_{\vec{k} + \lambda_{e} \vec{Q},\uparrow}^{\dagger} b_{\vec{k} - \lambda_{h} \vec{Q},\uparrow}|\psi_{o} \rangle  \label{eq:excitonb}
\end{equation}
The final state consists of no exciton and one photon spontaneously emitted with wavevector $-\vec{q}$ (where $\vec{q} = \vec{q}_{\parallel} + \hat{z} q_{z})$. Using Fermi's golden rule, summing over all possible final states, and assuming a finite phenomenological broadening (due to exciton scattering and/or polarization dephasing), we get for the spontaneous emission lifetime of the exciton,
\begin{eqnarray}
\frac{1}{\tau_{sp}(\vec{Q})} & = &  \frac{2\pi}{\hbar}  \sum_{\vec{q},j} \nonumber \\
& & \times \left\{ \left| \langle \psi_{o}|\langle - \vec{q},j|  H^{-}_{j}(\vec{q}) |0\rangle| \psi_{\vec{Q}}\rangle_{ex} \right|^{2}\right\}_{th} \nonumber \\
& & \times \frac{\Gamma_{ex}/\pi}{(E_{ex}(\vec{Q}) - \hbar \omega_q)^{2} + \Gamma_{ex}^{2}} \label{eq:ex_tau}
\end{eqnarray}    
The matrix element involving $H^{-}_{j}(\vec{q})$ will be zero unless $-\vec{q}_{\parallel}=\vec{Q}$, due to in-plane momentum conservation. The final result, after summing over the two possible polarizations of the emitted photon, can be written as,
\begin{eqnarray}
& & \frac{1}{\tau_{sp}(\vec{Q})} = \eta_{o} \frac{e^{2}}{m_{o}^{2}} |\chi_{ex}(\vec{r}=0,\vec{Q})|^{2} \int^{\infty}_{0} dq_{z} \frac{1}{\sqrt{Q^{2} + q_{z}^{2}}} \nonumber \\
& & \times \left(1 + \frac{q_{z}^{2}}{q_{z}^{2} + Q^{2}} \right)  \frac{\Gamma_{ex}/\pi}{(E_{ex}(\vec{Q}) - \hbar \sqrt{Q^{2} + q_{z}^{2}} c)^{2} + \Gamma_{ex}^{2}} \label{eq:ex_tau_2}
\end{eqnarray}
Here, $\eta_{o}$ is the free-space impedance, $c$ is the speed of light, and the function $\chi_{ex}(\vec{r},\vec{Q})$ is given as~\cite{Changjian14},
\begin{eqnarray}
  \chi_{ex}(\vec{r},\vec{Q}) & = & \int \frac{d^{2}\vec{k}}{(2\pi)^{2}} \vec{P}_{vc}(\vec{k}-\lambda_{h}\vec{Q},\vec{k}+\lambda_{e}\vec{Q}).\hat{x} \, \, \psi_{\vec{Q}}(\vec{k}) \nonumber \\
  & & \times \sqrt{1-f_{c}(\vec{k}+\lambda_{e}\vec{Q})} e^{i\vec{k}.\vec{r}} \label{eq:chi}
\end{eqnarray}
Any in-plane unit vector, other than $\hat{x}$, can be used in the expression above and the angle-averaged momentum matrix element will not depend on the specific choice. $\chi_{ex}(\vec{r},\vec{Q})$ incorporates the reduction in the exciton oscillator strength due to Pauli-blocking and band-filling effects~\cite{Changjian14}. Note that the phase factors in the definition of $\psi_{\vec{Q}}(\vec{k})$ (see (\ref{eq:psiphi})) cancel out exactly near the band edge in the expression for $\chi_{ex}(\vec{r},\vec{Q})$~\cite{Changjian14}. Equation (\ref{eq:ex_tau}) shows that only excitons with momenta $\vec{Q}$ that satisfy,
\begin{equation}
\hbar Q c < E_{ex}(\vec{Q}) = E_{g} - E_{exb} + \frac{\hbar^{2} Q^{2}}{2 m_{ex}} 
\end{equation}
radiate efficiently, as expected for excitons in 2D~\cite{Citrin93}. This implies that excitons in MoS$_{2}$ with kinetic energy greater than approximately $(E_{g}-E_{exb})^2/(2m_{ex}c^{2}) \sim 3.5$ $\mu$eV will not radiate efficiently unless assisted by phonons or impurities to provide in-plane momentum conservation. Since the exciton radius in MoS$_{2}$ is small, in the 7-10$\AA$ range~\cite{Changjian14}, accurate energy band dispersions and momentum matrix elements are needed for large wavevectors (at least a few nm$^{-1}$) in order to accurately describe exciton oscillator strengths~\cite{Changjian14}. Following Zhang et~al.~~\cite{Changjian14}, we use the Lowdin approximation to a 4-band model proposed by Kormanyos et~al.~\cite{Falko13}. This procedure adds the following matrix to the Hamiltonian given earlier in (\ref{eq:H1}), 
\begin{equation}
\left[
\begin{array}{cc}
\alpha k^{2} & \kappa k_{+}^2 - \frac{\eta}{2}k^{2}k_{-}    \\
\kappa k_{-}^{2} - \frac{\eta}{2}k^{2}k_{+} & \beta k^{2}
\end{array} \right]    \label{eq:H2}
\end{equation} 
\begin{figure}
  \begin{center}
   \epsfig{file=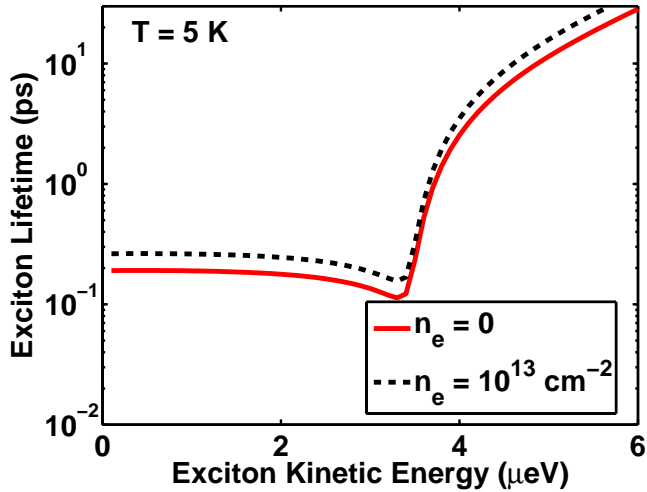,angle=0,width=0.48\textwidth}
    \caption{The radiative lifetime of excitons in suspended MoS$_{2}$ monolayer is plotted as a function of the in-plane exciton kinetic energy $\hbar Q^{2}/2m_{ex}$ for different doping densities $n_{e}$ (n-doped sample). Zero in-plane momentum excitons have extremely short radiative lifetimes in the 0.18-0.30 ps range as a result of the large optical oscillator strength of excitons in MoS$_{2}$ monolayer.}
    \label{fig:ex_lifetime}
  \end{center}
\end{figure} 

\begin{figure}
  \begin{center}
   \epsfig{file=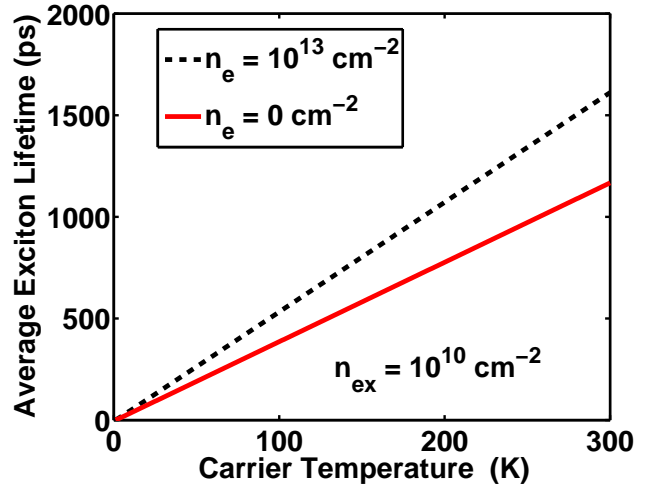,angle=0,width=0.48\textwidth}
    \caption{The average radiative lifetime of thermally distributed excitons is shown for different doping densities $n_{e}$ (n-doped sample) as a function of the carrier temperature (assumed to be the same for excitons and unbound electrons). The exciton density is assumed to be $10^{10}$ 1/cm$^{2}$. The average lifetime increases linearly with the temperature.}
    \label{fig:ex_lifetime_2}
  \end{center}
\end{figure} 
The values of the parameters $\alpha$, $\beta$, $\kappa$ and $\eta$ that best fit density functional theory (DFT) results are 1.72 eV$\AA^{2}$, -0.13  eV$\AA^{2}$, -1.02  eV$\AA^{2}$, and 8.52  eV$\AA^{3}$, respectively \cite{Falko13}. The resulting angle-averaged wavevector-dependent interband momentum matrix elements were given by Zhang et~al.~\cite{Changjian14}. In numerical calculations, we first compute exciton wavefunctions and radii as a function of the doping for different $\vec{Q}$ using a variational technique~\cite{Changjian14}. The procedure is iterated a few times to ensure that the optimal values of $\lambda_{e \atop h}$ are used. The wavevector-dependent momentum matrix elements are then used to obtain the exciton radiative lifetimes. Figure \ref{fig:ex_lifetime} shows the radiative lifetime of excitons as a function of the in-plane exciton kinetic energy $\hbar Q^{2}/2m_{ex}$ for different doping densities, $0$ and 10$^{13}$ 1/cm$^{2}$, and assuming $\Gamma_{ex} \approx 20$ meV. Excitons with small kinetic energy have extremely short radiative lifetimes in the 0.18-0.30 ps range. Such short radiative lifetimes are due to the small exciton radii and, therefore, large exciton optical oscillator strengths in MoS$_{2}$~\cite{Changjian14}. Pauli-blocking, coming directly from the factor $\sqrt{1-f_{c}}$ in Equation (\ref{eq:chi}), reduces the exciton oscillator strength when the doping increases. However, the exciton radius also decreases when the doping increases, as shown by Zhang et~al.~\cite{Changjian14}. The combined effect is that the exciton oscillator strength and the radiative lifetime do not change drastically with doping.

In most experimental situations, the radiative lifetime of an ensemble of excitons is measured. An ensemble can be created in different ways and could include both bright and dark excitons. We assume a thermal ensemble of only bright excitons. The procedure used to obtain the average lifetime of the ensemble is described in Appendix~\ref{app2}). Figure \ref{fig:ex_lifetime_2} shows the average radiative lifetime of thermally distributed excitons for different doping densities $n_{e}$ (n-doped sample) as a function of the carrier temperature (assumed to be the same for excitons and unbound electrons). The exciton density is assumed to be 10$^{10}$ 1/cm$^{2}$. The average lifetime increases linearly with the temperature and is largely independent of the exciton density (as long as the temperature is large enough, and the exciton density is small enough, such that the exciton chemical potential is several $KT$ smaller than the lowest exciton energy level). The average exciton radiative lifetime $\langle \tau_{sp} \rangle$ can be related to the zero-momentum lifetime  $\tau_{sp} (\vec{Q}=0) $ by,
\begin{equation}
\langle \tau_{sp} \rangle \approx \frac{KT}{E_{o}} \tau_{sp} (\vec{Q}=0) \;\;\;\;\;\; {\rm for} \; KT > E_{o} \label{eq:ensemble}
\end{equation}
where, $E_{o}$, given approximately by $(E_{g}-E_{exb})^{2}/2m_{ex}c^{2}$, is the kinetic energy of the largest in-plane momentum exciton that can radiate. Note that the value of $\Gamma_{ex}$ does not affect the result for $\tau_{sp}(\vec{Q}=0)$ in any significant way and nor does it affect the average exciton radiative lifetime of a thermal ensemble.

\subsection{Transverse and Longitudinal Excitons}
Superpositions of exciton states from the two valleys can be created such that these exciton states couple to only the transverse (TE) or the longitudinal (TM) polarized radiation fields. These longitudinal and transverse exciton states, and their radiative lifetimes, are discussed in Appendix~\ref{app1} where we also discuss the dipole-dipole exchange interactions between excitons.

\section{Photon Emission by Trions} \label{sec:trions}

\subsection{Singlet Trion States with Non-Zero Momentum}
A trion is formed when a photoexcited electron-hole pair binds with an electron (or a hole) to form a negatively (or a positively) charged complex. A trion state can be bright or dark and can be a spin singlet or a triplet. In this paper we consider only bright singlet trions~\cite{Changjian14}. Without losing generality, we restrict ourselves to negatively charged trions (relevant to n-doped samples). We also restrict ourselves to the case $\tau=1$ where the top most valence band is occupied by spin-up ($\sigma=1$) electrons, and drop the spin-valley index. We define the trion mass as $m_{tr} = 2m_{e} + m_{h}$. As before, we assume that the ground state $|\psi_{o}\rangle$ of the semiconductor consists of a completely filled valence band and a conduction band with an electron density $n_{e}$ distributed according to the Fermi-Dirac distribution. A bright singlet trion state with momentum $\vec{Q}$ can be constructed from the ground state as follows,
\begin{eqnarray}
|\psi_{\vec{Q}}\rangle_{tr}  & = &  \frac{1}{A}\sum_{\vec{k}_{1},\vec{k}_{2}} \frac{\psi_{\vec{Q}}(\vec{k}_{1},\vec{k}_{2})} {N_{\vec{Q}}(\vec{k_{1}},\vec{k_{2}})}  \times \nonumber \\
&& c_{\ul{\vec{k}_{1}}, \downarrow}^{\dagger}  c_{\ul{\vec{k}_{2}}, \uparrow}^{\dagger} b_{\vec{k}_{1} + \vec{k}_{2} - \eta_{h} \vec{Q}, \uparrow}  |\psi_{o} \rangle   \label{eq:trion}
\end{eqnarray}
Here, the line under a vector, $\ul{\vec{k}}$, stands for $\vec{k}+ \eta_{e} \vec{Q}$. $\eta_{e}$ and $\eta_{h}$ equal $m_{e}/m_{tr}$ and $m_{h}/m_{tr}$, respectively. As in the case of the exciton, the average values of the inverse electron and hole effective masses with respect to the trion relative wavefunction are used in these expressions. The trion wavefunction $\psi_{\vec{Q}}(\vec{k}_{1},\vec{k}_{2})$ is symmetric in its first two arguments. The normalization factor $N_{\vec{Q}}(\vec{k_{1}},\vec{k_{2}})$ equals,
\begin{equation}
\sqrt{(1 - f_{c}(\ul{\vec{k}_{2}})) (1 - f_{c}(\ul{\vec{k}_{1}}))}
\end{equation}
The trion wavefunction is normalized such that $\left\{ _{tr}\langle \psi_{\vec{Q}}|\psi_{\vec{Q}} \rangle_{tr} \right\}_{th} =1$, where the curly brackets represent averaging with respect to the thermal ensemble. This normalization gives, 
\begin{equation}
\int \frac{d^{2}\vec{k_{1}}}{(2\pi)^{2}} \frac{d^{2}\vec{k_{2}}}{(2\pi)^{2}} \psi^{*}_{\vec{Q}}(\vec{k}_{1},\vec{k}_{2}) \psi_{\vec{Q}}(\vec{k}_{1},\vec{k}_{2}) = 1
\end{equation}
If one writes the trion wavefunction as (including the valley index $\tau$),
\begin{equation}
\psi_{\vec{Q},\tau}(\vec{k}_{1},\vec{k}_{2})  = \phi_{\vec{Q}}(\vec{k}_{1},\vec{k}_{2}) e^{i(\tau\phi_{\ul{\vec{k}_{1}}} + \tau\phi_{\ul{\vec{k}_{2}}} + \tau\phi_{\vec{k}_{1} + \vec{k}_{2} - \eta_{h} \vec{Q}})/2} \label{eq:psiphi2}
\end{equation}
then, assuming (\ref{eq:F2}), and using the trion state given in (\ref{eq:trion}) as a variational state, the trion wavefunction $\phi_{\vec{Q}}(\vec{k}_{1},\vec{k}_{2})$ satisfies the Hermitian eigenvalue equation~\cite{Changjian14},
\begin{eqnarray}
&& \left[ \bar{E}_{c}(\ul{\vec{k_{1}}}) + \bar{E}_{c}(\ul{\vec{k_{2}}})  \right. \nonumber \\
&& \left. - \bar{E}_{v}(\vec{k_{1}}+\vec{k_{2}}-\eta_{h}\vec{Q}) \right]\phi_{\vec{Q}}(\vec{k}_{1},\vec{k}_{2}) \nonumber \\
&&  + \frac{\sqrt{1 - f_{c}(\ul{\vec{k}_{1}})} \sqrt{1 - f_{c}(\ul{\vec{k}_{2}})}}{A} \sum_{\vec{q}} \left[ V(\vec{q}) \phi_{\vec{Q}}(\vec{k_{1}}-\vec{q},\vec{k_{2}}+\vec{q}) \right. \nonumber \\
&& \left. \sqrt{1 - f_{c}(\ul{\vec{k}_{1}}-\vec{q})} \sqrt{1 - f_{c}(\ul{\vec{k}_{2}}+\vec{q})}  \right] \nonumber \\
&& - \frac{\sqrt{1 - f_{c}(\ul{\vec{k}_{1}})} }{A} \sum_{\vec{q}} \left[ V(\vec{q})\, \phi_{\vec{Q}}(\vec{k_{1}}-\vec{q},\vec{k_{2}}) \right. \nonumber \\
&& \left. \sqrt{1 - f_{c}(\ul{\vec{k}_{1}}-\vec{q})} \right] - \frac{\sqrt{1 - f_{c}(\ul{\vec{k}_{2}})} }{A} \sum_{\vec{q}} \left[ V(\vec{q})\, \phi_{\vec{Q}}(\vec{k_{1}},\vec{k_{2}-\vec{q}}) \right. \nonumber \\
&& \left. \sqrt{1 - f_{c}(\ul{\vec{k}_{2}}-\vec{q})} \right] = E_{tr}(\vec{Q}) \phi_{\vec{Q}}(\vec{k}_{1},\vec{k}_{2})  \label{eq:eigen2}
\end{eqnarray}
The trion energy $E_{tr}(\vec{Q})$ is measured with respect to the energy of the ground state $|\psi_{o}\rangle$. {\em Note that all the phase factors cancel out and do not appear in the trion eigenvalue equation}. For small trion momentum $\vec{Q}$, the band energy dispersions can be Taylor expanded and $E_{tr}(\vec{Q})$ can be expressed as~\cite{Changjian14},
\begin{equation}
E_{tr}(\vec{Q}) =  E_{g} - E_{exb} - E_{trb} + \frac{\hbar^{2}Q^{2}}{2m_{tr}} 
\end{equation}
where $E_{trb}$ is the trion binding energy.   

 \subsection{Radiative Lifetime of Trions}
\begin{figure}
  \begin{center}
   \epsfig{file=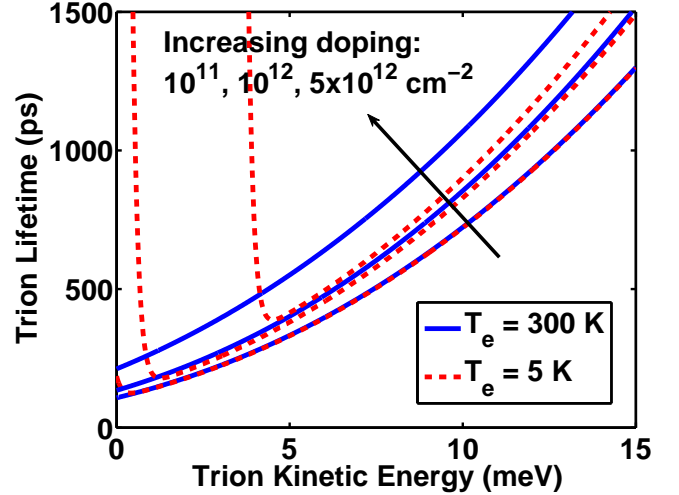,angle=0,width=0.48\textwidth}
    \caption{The radiative lifetime of trions in suspended MoS$_{2}$ monolayer is plotted as a function of the in-plane trion kinetic energy $\hbar Q^{2}/2m_{tr}$ for different doping densities ($10^{11}$, $10^{12}$ and $5\times 10^{12}$ 1/cm$^{2}$), and two different electron temperatures, 5 K and 300 K.}
    \label{fig:tr_lifetime}
  \end{center}
\end{figure}

\begin{figure}
  \begin{center}
   \epsfig{file=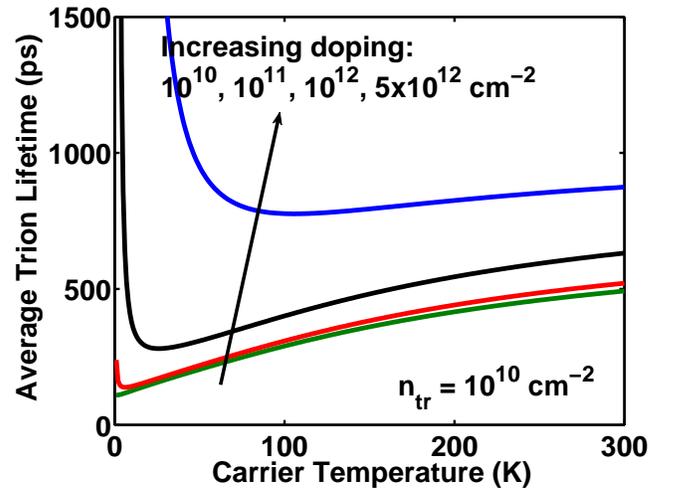,angle=0,width=0.48\textwidth}
    \caption{The average radiative lifetime of thermally distributed trions is shown for different doping densities $n_{e}$ (n-doped sample) as a function of the carrier temperature (assumed to be the same for trions and unbound electrons). The trion density $n_{tr}$ is assumed to be $10^{10}$ 1/cm$^{2}$. As the temperature increases from zero, the average trion lifetime first decreases, reaches a minimum value, and then increases.}
    \label{fig:tr_lifetime_2}
  \end{center}
\end{figure}

The interaction Hamiltonian given in Section \ref{sec:int} can be used for trions as well. We assume that the initial state is as given in (\ref{eq:trion}). The final state consists of no trion, one photon spontaneously emitted with momentum $-\vec{q}$ ($\vec{q} = \vec{q}_{\parallel} + \hat{z} q_{z}$), and an electron left in the conduction band that carries the momentum $\vec{Q} + \vec{q}_{\parallel}$. Using Fermi's golden rule, summing over all possible final states, and assuming a finite broadening, we get for the spontaneous emission lifetime of a trion,
\begin{eqnarray}
& & \frac{1}{\tau_{sp}(\vec{Q})} = \frac{\pi e^{2}}{\epsilon_{o} m_{o}^{2}} \int \frac{d^{3}\vec{q}}{(2\pi)^{3}} \frac{1}{\omega_q} \left(1 + \frac{q_{z}^{2}}{q^{2}} \right) \times \nonumber \\
& & \left| \int d^{2}\vec{r}_{1} \chi_{tr}(\vec{r}_{1}, \vec{r}_{2}=0, \vec{q}_{\parallel},\vec{Q}) e^{-i[\vec{q}_{\parallel} + (m_{ex}/m_{tr})\vec{Q}].\vec{r}_{1}}   \right|^{2} \times \nonumber \\
& & \left[ 1-f_{c}(\vec{Q}+\vec{q}_{\parallel}) \right] \frac{\Gamma_{tr}/\pi}{ \left( E_{tr}(\vec{Q})  - \frac{\hbar^{2}|\vec{Q}+\vec{q}_{\parallel}|^{2}}{2m_{e}} - \hbar \omega_q \right)^{2} + \Gamma_{tr}^{2}} \label{eq:tr_time}
\end{eqnarray}
where,
\begin{eqnarray}
& & \chi_{tr}(\vec{r_{1}},\vec{r_{2}},\vec{q}_{\parallel},\vec{Q}) =  \int \frac{d^{2}\vec{k_{1}}}{(2\pi)^{2}} \int \frac{d^{2}\vec{k_{2}}}{(2\pi)^{2}} \vec{P}_{vc}(\ul{\vec{k_{2}}} + \vec{q}_{\parallel}, \ul{\vec{k_{2}}} ).\hat{x}  \nonumber \\
  & & \times e^{-i\tau\phi_{\ul{\vec{k}_{1}}}/2} \psi_{\vec{Q}}(\vec{k_{1}},\vec{k_{2}}) \sqrt{1-f_{c}(\ul{\vec{k_{2}}})} \, \, e^{i\vec{k_{1}}.\vec{r_{1}} +i\vec{k_{2}}.\vec{r_{2}} }
\end{eqnarray}
Again note that any in-plane unit vector, other than $\hat{x}$, can be used in the expression above and the angle-averaged momentum matrix element will not depend on the specific choice. $\chi_{tr}(\vec{r_{1}},\vec{r_{2}},\vec{q}_{\parallel},\vec{Q})$ incorporates the reduction in the trion oscillator strength due to Pauli-blocking and band-filling effects~\cite{Changjian14}. Note that the phase factors in the definition of $\psi_{\vec{Q}}(\vec{k_{1}},\vec{k_{2}})$ (see (\ref{eq:psiphi2})) cancel out exactly near the band edge in the expression above for the trion lifetime~\cite{Changjian14}. In calculations, we first compute the trion wavefunctions and radii as a function of the doping for different $\vec{Q}$ using a variational technique~\cite{Changjian14}, and then use the wavevector dependent momentum matrix elements to obtain the trion radiative lifetimes. Figure \ref{fig:tr_lifetime} shows the radiative lifetime of trions as a function of the in-plane trion kinetic energy $\hbar Q^{2}/2m_{tr}$ for different doping densities ($10^{11}$, $10^{12}$ and $5\times 10^{12}$ 1/cm$^{2}$), and two different electron temperatures, 5 K and 300 K. The trion lifetime increases with the trion kinetic energy because the squared matrix element involving $\chi_{tr}$ in (\ref{eq:tr_time}) decreases with the increase in the magnitude of $\vec{Q}$. The longer trion lifetime seen for small trion kinetic energies at low temperatures and large doping densities is due to the fact that the phase space for the electron, which got left behind in the conduction band when the trion recombined, is either not available or is much reduced near the conduction band edge because of Pauli-blocking. Figure \ref{fig:tr_lifetime_2} shows the average radiative lifetime of thermally distributed trions for different doping densities $n_{e}$ (n-doped sample) as a function of the carrier temperature (assumed to be the same for trions and unbound electrons). The trion density $n_{tr}$ is assumed to be $10^{10}$ 1/cm$^{2}$ and consists of only bright singlet trions. As the temperature increases from zero, the average trion lifetime first decreases, reaches a minimum value, and then increases. The initial decrease in the lifetime with the temperature occurs because the phase space for the left-behind electron near the conduction band bottom opens up as the temperature increases and the electron density (from the doping) shifts to higher energies. In lightly doped samples this initial decrease occurs at very small temperatures and might not be observable in experiments. As the temperature increases further, the trions acquire more thermal energy and their distribution spreads to higher energies. Since trions with large kinetic energies have longer lifetimes (as shown in Figure \ref{fig:tr_lifetime}), the average trion lifetime increases with the temperature. The increase in the average trion lifetime with the temperature is consistent with the experimental results for trions in III-V semiconductor quantum wells~\cite{Pepper00}.

\section{Effects of Localization}

\begin{figure}
  \begin{center}
   \epsfig{file=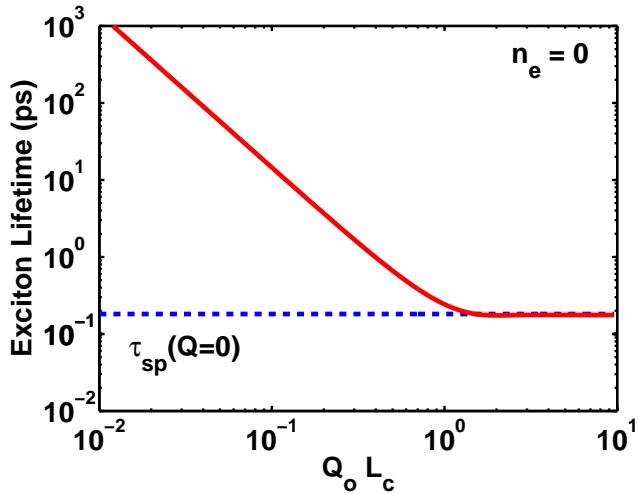,angle=0,width=0.48\textwidth}
    \caption{The radiative lifetime of a localized exciton in suspended MoS$_{2}$ monolayer is plotted as a function of the localization length $L_{c}$. $Q_{o}$ is the momentum of a photon of energy equal to the exciton energy.  In a MoS$_{2}$ monolayer, $Q_{o}\sim  10^{7} 1/m$. The dashed line represents, for reference, the radiative lifetime of a free (unlocalized) exciton with zero in-plane momentum $\vec{Q}$. The doping density $n_{e}$ is assumed to be zero.}
    \label{fig:ex_lifetime_loc}
  \end{center}
\end{figure} 

\subsection{Radiative Lifetime of Localized Excitons}
Since radiative rates of excitons are limited by momentum conservation requirements, localization can enhance or reduce radiative rates of excitons by broadening the distribution of the center of mass momenta of the exciton states~\cite{Citrin93}. Following Citrin et~al.~\cite{Citrin93}, we consider excitons localized in space in regions of size $L_{c}$. We assume that the exciton wavefunction for the center of mass coordinate in real and Fourier spaces is,
\begin{eqnarray} 
  & & \psi_{com}(\vec{R}) = \frac{1}{\sqrt{\pi L_{c}^{2}}} e^{-R^{2}/2L_{c}^{2}} \nonumber \\
  & & \psi_{com}(\vec{Q}) = \sqrt{4\pi L_{c}^{2}} e^{-Q^{2}L_{c}^{2}/2} 
\end{eqnarray}
We again restrict ourselves to the case $\tau=1$ where the top most valence band is occupied by spin-up ($\sigma=1$) electrons, and drop the spin-valley index. A localized exciton state can be constructed from the ground state $|\psi_{o} \rangle $ using a superposition of exciton states of different momenta,
\begin{eqnarray}
|\psi_{ex} \rangle & & =  \frac{1}{A} \sum_{\vec{Q}} \psi_{com}(\vec{Q}) \sum_{\vec{k}} \frac{\psi_{\vec{Q}}(\vec{k})}{N_{\vec{Q}}(\vec{k})} \times \nonumber \\
& & c_{\vec{k} + \lambda_{e}\vec{Q}, \uparrow}^{\dagger} b_{\vec{k}- \lambda_{h} \vec{Q}, \uparrow}|\psi_{o} \rangle  \label{eq:exciton2}
\end{eqnarray}
Assuming that the kinetic energy associated with the center of mass coordinate is $E_{com}$, the energy of the exciton relative to the ground state $|\psi_{o} \rangle $ is,
\begin{equation}
E_{ex} = E_{g} - E_{exb} + E_{com} = \hbar Q_{o} c
\end{equation}
Here, we have defined $Q_{o}$ as the momentum of a photon of energy equal to the exciton energy. As discussed below, the product $Q_{o}L_{c}$ will govern the exciton radiative lifetime. Ignoring, for simplicity, the $\vec{Q}$ dependence of the function $\chi_{ex}(\vec{r},\vec{Q})$ ($\chi_{ex}(\vec{r},\vec{Q}) \approx \chi_{ex}(\vec{r},\vec{Q}=0)$ since the $\vec{Q}$ dependence of $\chi_{ex}(\vec{r},\vec{Q})$ is weak and only excitons with very small $\vec{Q}$ radiate) the spontaneous emission lifetime of the localized exciton is, 
\begin{eqnarray}
  & & \frac{1}{\tau_{sp}(L_{c})} = \frac{\pi e^{2}}{\epsilon_{o} m_{o}^{2}} |\chi_{ex}(\vec{r}=0,\vec{Q}=0)|^{2} \nonumber \\
  & & \times \int \frac{d^{3}\vec{q}}{(2\pi)^{3}} \frac{|\psi_{com}(\vec{q}_{\parallel})|^{2}}{\omega_{q}} \nonumber \\
& & \times \left(1 + \frac{q_{z}^{2}}{q^{2}} \right)  \frac{\Gamma_{ex}/\pi}{(E_{ex}(\vec{Q} = \vec{q}_{\parallel}) - \hbar \omega_{q})^{2} + \Gamma_{ex}^{2}}
\end{eqnarray}
Figure \ref{fig:ex_lifetime_loc} shows the radiative lifetime of a localized exciton in a suspended MoS$_{2}$ monolayer as a function of the localization length $L_{c}$. When $L_{c} << Q_{o}^{-1}$, the exciton wavefunction spread in momentum space is larger than $Q_{o}$ and the exciton does not radiate efficiently. In this regime, the exciton lifetime increases with the decrease in the localization length as,
\begin{equation}
\tau_{sp}(L_{c}) \approx  \frac{\tau_{sp}(\vec{Q}=0)}{(Q_{o}L_{c})^{2}} \;\;\;\;\; {\rm for} \; \; Q_{o}L_{c}\ll 1
\end{equation}
When $L_{c} >> Q_{o}^{-1}$, the exciton wavefunction in momentum space is localized in a region smaller than $Q_{o}$ and the exciton lifetime is essentially the same as that of a free (unlocalized) exciton with near-zero momentum. The results in Figure \ref{fig:ex_lifetime_loc} are quantitatively valid only if $L_{c}$ is larger than the exciton radius ($\sim$1 nm) otherwise the relative wavefunction of the exciton state would also change as a result of localization - an effect we have ignored.

In a MoS$_{2}$ monolayer, $Q_{o}\sim  10^{7} 1/m$. This means that the relevant length scale against which to compare $L_{c}$ is $\sim$100 nm. The temperature dependence of the exciton radiative lifetime measured in experiments will depend on the nature of localization. For example, if all the excitons are strongly localized, and remain localized as the temperature is increased, then no particular temperature dependence of the average radiative lifetime is expected~\cite{Tomimoto07}. The short lifetimes of weakly localized excitons in Figure \ref{fig:ex_lifetime_loc} for $Q_{o} L_{c} > 0.1$ (or $L_{c} > 10$ nm in MoS$_{2}$), may not be observed in experiments as the energy level spacings in such large localizing regions are likely to be small and the excitons could occupy higher energy levels after gaining thermal energy and their ensemble averaged lifetime would then exhibit the characteristics of free excitons.

\section{Discussion}
In this paper, we presented radiative lifetimes of excitons and trions in MoS$_{2}$ monolayers. The analysis presented here is also applicable to excitons and trions in other metal dichalcogenides since the optical material properties of sulfides and selenides of molybdenum and tungsten relevant to the radiative lifetimes of excitons and trions are not too different. Recent experimental results on the lifetimes of near-zero momentum excitons in WSe$_{2}$ monolayer have yielded numbers in the 150-250 fs range~\cite{Huber15,Moody15,Marie15} in good agreement with the numbers calculated in this paper for MoS$_{2}$ (Figure  \ref{fig:ex_lifetime}). In addition, very long ($\ge$1 ns) average exciton lifetimes have also been observed in clean MoS$_{2}$ monolayers at very small photoexcited exciton densities~\cite{Amani15}. These long lifetime values are consistent with strongly localized excitons (Figure \ref{fig:ex_lifetime_loc}).

Some caution must be exercised in interpreting the results presented in this paper. First, given the extremely short radiative lifetimes of near-zero momentum free excitons (Figure \ref{fig:ex_lifetime}), the assumption of a thermal ensemble for free excitons might not be relevant to experiments. It is likely that the average radiative lifetimes of free excitons observed in experiments are determined by the energy relaxation rates of excitons or by assisted processes, such as phonon and impurity scattering, which would provide the momentum necessary for excitons with large in-plane momentum to radiate. Second, in real samples the excitons might become localized at low temperatures. In this case, as the temperature is lowered, the observed average radiative lifetime of the excitons would decrease linearly with the temperature at high temperatures and become weakly dependent on the temperature at low temperatures. Third, our analysis of the trion radiative lifetimes did not take into account the fact that trions are more likely to form in the first place when the doping density is large and, therefore, experimentally observed trion photoluminescence quantum efficiencies at different doping densities might not reflect the trends in trion lifetimes with doping depicted in Figures \ref{fig:tr_lifetime} and \ref{fig:tr_lifetime_2}. In fact, a better understanding of the formation dynamics of both excitons and trions would be needed in addition to the results presented here in order to correctly interpret photoluminescence data~\cite{Kira06}.

\section{Acknowledgments}
The authors would like to acknowledge helpful discussions with Jared Strait, Paul L. McEuen and Michael G. Spencer, and support from CCMR under NSF grant number DMR-1120296, AFOSR-MURI under grant number FA9550-09-1-0705, and ONR under grant number N00014-12-1-0072.

\section{Appendices}

\subsection{Exciton Self-Energy, Exchange Interactions, and Radiative Lifetimes: A Green's Function Approach} \label{app1}
In this section, we calculate the exciton self-energy in TMD monolayers. The exciton eigenstates in the presence of long-range exciton dipole-dipole exchange interactions are longitudinal and transverse excitons. We show that the exciton dispersion for small momenta (inside the light cone) is not modified as a result of these interactions.

We assume an undoped TMD monolayer for simplicity. We consider a reduced Hilbert space in which only the lowest exciton state is relevant. The exciton creation operator is~\cite{Wang15b},
\begin{equation}
  B^{\dagger}_{\vec{Q},s} =  \frac{1}{\sqrt{A}}\sum_{\vec{k}} \psi_{\vec{Q},s}(\vec{k}) c_{\vec{k} + \lambda_{e} \vec{Q},s}^{\dagger} b_{\vec{k}-\lambda_{h}\vec{Q},s} 
\end{equation}
Products of electron and hole creation and destruction operators can thus be expressed in terms of the exciton operators. We define an exciton polarization vector as,
\begin{eqnarray}
  \hat{m}_{\vec{Q},s}  & = & \frac{|\chi_{ex}(0,\vec{Q})|^{-1}}{\sqrt{2}A} \sum_{\vec{k}} \vec{P}_{vc,s}(\vec{k} - \lambda_{h}\vec{Q}, \vec{k} + \lambda_{e}\vec{Q}) \psi_{\vec{Q},s}(\vec{k}) \nonumber \\
\end{eqnarray}
with $\hat{m}^{*}_{\vec{Q},s} \cdotp \hat{m}_{\vec{Q},s} =1$. It is convenient to define the exciton vector field operator as,
\begin{equation}
\vec{C}_{\vec{Q}}(t) =  \sum_{s} \hat{m}_{\vec{Q},s} B_{\vec{Q},s}(t) + \hat{m}^{*}_{-\vec{Q},s} B^{\dagger}_{-\vec{Q},s}(t)
\end{equation}
$\vec{C}_{\vec{Q}}(t)$ is proportional to the Fourier component of the exciton interband polarization at wavevector $\vec{Q}$. The non-interacting retarded Green's function for the exciton field is~\cite{Mahan00},
\begin{equation}
  \overline{\overline{G}}^{oR}_{\vec{Q}}(t-t') = -\frac{i}{\hbar}\theta(t-t')\langle[\vec{C}_{\vec{Q}}(t),\vec{C}_{-\vec{Q}}(t')]\rangle
  \end{equation}
and in the Fourier domain it equals,
\begin{eqnarray}
  \overline{\overline{G}}^{oR}_{\vec{Q}}(\omega) & \approx & \frac{2E_{ex}(\vec{Q})}{(\hbar\omega)^{2}-E_{ex}(\vec{Q})^{2} + i\eta} \left[ \hat{x}\otimes \hat{x} + \hat{y}\otimes \hat{y} \right] \nonumber \\
  & = & \frac{2E_{ex}(\vec{Q})}{(\hbar\omega)^{2}-E_{ex}(\vec{Q})^{2} + i\eta} \left[ \begin{array}{cc} 1 & 0 \\ 0 & 1 \end{array} \right]
\end{eqnarray}
Note that summation over the spin-valley index makes the Green's function matrix conveniently diagonal.  

The quantized radiation can be expressed in terms of the following field (which is proportional to the Fourier component of the vector potential),
\begin{equation}
\vec{R}_{\vec{q}}(t) = \sum_{j} \hat{n}_{j}(\vec{q}) a_{j}(\vec{q},t) + \hat{n}^{*}_{j}(-\vec{q})  a^{\dagger}_{j}(-\vec{q},t) 
\end{equation}
The interaction Hamiltonian given earlier in (\ref{eq:light_H2}) can be written as,
\begin{equation}
  \hat{H}_{int} =  \frac{e}{m_{o}} \sum_{\vec{q}}  \sqrt{\frac{A\hbar}{V \epsilon_{o} \omega_q}} |\chi_{ex}(0,\vec{q}_{\parallel})| \vec{R}_{\vec{q}} \,\, \cdotp \, \vec{C}_{-\vec{q}_{\parallel}} \label{eq:light_H3}
\end{equation}
Unlike in the main text of this paper where the goal was to compute the radiative lifetimes (a resonant linear optical process), here we are interested in the exciton self-energies, both on-shell and off-shell. The term quadratic in the vector potential must be added to the interaction Hamiltonian. Following Girlanda et~al.~\cite{Girlanda95}, the form of this extra term is found to be,
\begin{eqnarray}
  & & \hat{H}'_{int} =  \sum_{\vec{Q},s} E_{ex}(\vec{Q}) \nonumber \\
  & & \times \left[ \frac{e}{m_{o}} \sum_{\vec{q}} \delta_{\vec{q}_{\parallel},\vec{Q}} \sqrt{\frac{A\hbar}{V \epsilon_{o} \omega_q}} \frac{|\chi_{ex}(0,\vec{q}_{\parallel})|}{E_{ex}(\vec{q}_{\parallel})} \vec{R}_{\vec{q}} \,\, \cdotp \, \hat{m}^{*}_{\vec{q}_{\parallel},s} \right] \nonumber \\
  & & \times \left[ \frac{e}{m_{o}} \sum_{\vec{k}} \delta_{\vec{k}_{\parallel},\vec{Q}} \sqrt{\frac{A\hbar}{V \epsilon_{o} \omega_k}} \frac{|\chi_{ex}(0,\vec{k}_{\parallel})|}{E_{ex}(\vec{k}_{\parallel})} \vec{R}_{-\vec{k}} \,\, \cdotp \, \hat{m}_{\vec{k}_{\parallel},s} \right] \nonumber \\
  \end{eqnarray}
The non-interacting retarded Green's function for the radiation field is gauge-dependent. It is convenient to choose the temporal gauge in which the scalar potential is set equal to zero. In the temporal gauge the radiation Green's function is~\cite{Guad80},
 \begin{eqnarray}
\overline{\overline{D}}^{oR}_{\vec{q}}(t-t') & = & -\frac{i}{\hbar}\theta(t-t')\langle[\vec{R}(\vec{q},t),\vec{R}(-\vec{q},t')]\rangle \nonumber \\
\overline{\overline{D}}^{oR}_{\vec{q}}(\omega) & = &    \frac{2\hbar \omega_{q}}{(\hbar\omega)^{2} - (\hbar \omega_{q})^{2} + i\eta} \left[ 1 - \frac{\vec{q} \otimes \vec{q}}{\omega^{2}/c^{2}} \right] \nonumber \\
\end{eqnarray}
The projection of the Green's function in the plane of the monolayer is represented by,
\begin{eqnarray}
  \overline{\overline{D}}^{oR}_{\vec{q}}(\omega) & = &  \frac{2\hbar \omega_{q}}{(\hbar\omega)^{2} - (\hbar \omega_{q})^{2} + i\eta} \nonumber \\
  & & \times \left[ \begin{array}{cc} 1-\frac{q^{2}_{x}}{\omega^{2}/c^{2}} & -\frac{q_{x}q_{y}}{\omega^{2}/c^{2}} \\ -\frac{q_{x}q_{y}}{\omega^{2}/c^{2}} & 1 - \frac{q^{2}_{y}}{\omega^{2}/c^{2}} \end{array} \right] \nonumber \\
\end{eqnarray}
The Dyson equation for the exciton Green's function becomes,
\begin{equation}
  \overline{\overline{G}}^{R}_{\vec{Q}}(\omega) =   \overline{\overline{G}}^{oR}_{\vec{Q}}(\omega) +   \overline{\overline{G}}^{oR}_{\vec{Q}}(\omega) \, \cdotp \, \overline{\overline{\Sigma}}^{R}_{\vec{Q}}(\omega) \, \cdotp \, \overline{\overline{G}}^{R}_{\vec{Q}}(\omega)
  \end{equation}
where the retarded self-energy matrix is found to be,
\begin{eqnarray}
  \overline{\overline{\Sigma}}^{R}_{\vec{Q}}(\omega) & \approx & \left( \frac{e}{m_{o}} \right)^{2} |\chi_{ex}(0,\vec{Q})|^{2} \sum_{\vec{q}} \delta_{\vec{q}_{\parallel},\vec{Q}} \left( \frac{\hbar \omega}{E_{ex}(\vec{q}_{\parallel})} \right)^{2} \nonumber \\
  & & \times \left( \frac{A\hbar}{V \epsilon_{o} \omega_q} \right) \overline{\overline{D}}^{oR}_{\vec{q}}(\omega)
\end{eqnarray}
The self-energy matrix is not diagonal in the chosen coordinate basis. This can be fixed by rotating the coordinate basis such that the directions longitudinal ($L$) and transverse ($T$) to the in-plane vector $\vec{Q}$ are chosen as the basis. The exciton self-energy then becomes,
\begin{equation}
\overline{\overline{\Sigma}}^{R}_{\vec{Q}}(\omega) = \left[ \begin{array}{cc}  \Sigma^{R}_{\vec{Q},L}(\omega) & 0 \\ 0 &  \Sigma^{R}_{\vec{Q},T}(\omega) \end{array} \right] 
\end{equation}  
where,
\begin{eqnarray}
  & &  \Sigma^{R}_{\vec{Q},T}(\omega) = \left( \frac{e}{m_{o}} \right)^2 |\chi_{ex}(0,\vec{Q})|^{2} \sum_{\vec{q}} \delta_{\vec{q}_{\parallel},\vec{Q}} \left( \frac{\hbar \omega}{E_{ex}(\vec{q}_{\parallel})} \right)^{2} \nonumber \\
  & & \times \left( \frac{A\hbar}{V \epsilon_{o} \omega_q} \right) \frac{2\hbar \omega_{q}}{(\hbar\omega)^{2} - (\hbar \omega_{q})^{2} + i\eta}
  \end{eqnarray}
\begin{eqnarray}
  & &  \Sigma^{R}_{\vec{Q},L}(\omega) = \left( \frac{e}{m_{o}} \right)^2 |\chi_{ex}(0,\vec{Q})|^{2} \sum_{\vec{q}} \delta_{\vec{q}_{\parallel},\vec{Q}} \left( \frac{\hbar \omega}{E_{ex}(\vec{q}_{\parallel})} \right)^{2} \nonumber \\
  & & \times \left( \frac{A\hbar}{V \epsilon_{o} \omega_q} \right) \left( 1 - \frac{q^{2}_{\parallel}}{\omega^{2}/c^{2}} \right) \frac{2\hbar \omega_{q}}{(\hbar\omega)^{2} - (\hbar \omega_{q})^{2} + i\eta}
  \end{eqnarray}
The corresponding longitudinal and transverse Green's functions for the exciton are,
\begin{eqnarray}
  G^{R}_{\vec{Q},L/T}(\omega) =  \frac{2E_{ex}(\vec{Q})}{(\hbar\omega)^{2}-E_{ex}(\vec{Q})^{2} - 2E_{ex}(\vec{Q})\Sigma^{R}_{\vec{Q},L/T}(\omega)}
    \end{eqnarray}
The longitudinal and transverse excitons are the eigenstates in the presence of the exciton dipole-dipole exchange interaction. One can define creation operators for the longitudinal and the transverse excitons as follows,
\begin{equation}
B^{\dagger}_{\vec{Q},L/T} = \frac{1}{\sqrt{2}} \left[ B^{\dagger}_{\vec{Q},\tau=1} \pm e^{2i\phi_{\vec{Q}}} B^{\dagger}_{\vec{Q},\tau=-1} \right] 
\end{equation}
The dispersion of the longitudinal and the transverse excitons can be found from the poles of the corresponding Green's functions. When $\omega > Qc$ (inside the light cone), both the self-energies $\Sigma^{R}_{\vec{Q},L/T}(\omega)$ have a vanishingly small real part and a large magnitude of the imaginary part. The latter corresponds to the radiative lifetime of the exciton. The situation is reversed when $\omega < Qc$ and then the magnitude of the real part of the self-energy becomes large and the imaginary part vanishes. Therefore, when $E_{ex}(\vec{Q}) > \hbar Qc$ (inside the light cone) one can ignore corrections to the exciton dispersion from dipole-dipole interactions, and the lifetime of the exciton can be related to the imaginary part of the self-energy evaluated on the shell,
\begin{equation}
  \frac{1}{\tau_{sp,L/T}(\vec{Q})} = -\frac{2}{\hbar}{\rm Img}\left\{ \Sigma^{R}_{\vec{Q},L/T}(\omega) \right\}_{\hbar \omega = E_{ex}(\vec{Q})}
\end{equation}
and, we get,
\begin{eqnarray}
  \frac{1}{\tau_{sp,T}(\vec{Q})} & = & \frac{ 2\eta_{o} e^{2}}{m^{2}_{o}} |\chi_{ex}(0,\vec{Q})|^{2}\frac{1}{\sqrt{E^{2}_{ex}(\vec{Q}) - (\hbar Q c)^{2}}} \nonumber \\
  \frac{1}{\tau_{sp,L}(\vec{Q})} & = &  \frac{ 2\eta_{o} e^{2}}{m^{2}_{o}} |\chi_{ex}(0,\vec{Q})|^{2}\frac{\sqrt{E^{2}_{ex}(\vec{Q}) - (\hbar Q c)^{2}}}{E^{2}_{ex}(\vec{Q})} \nonumber \\
\end{eqnarray}
The expression for the radiative rate given in the main text in (\ref{eq:ex_tau_2}), in the limit $\Gamma_{ex} \rightarrow 0$, is exactly equal to the average of the longitudinal and transverse radiative rates given above. This is because an exciton state belonging to only one valley is an equal superposition state of longitudinal and transverse excitons. When $\omega < Qc$ and the self-energies are real, we get,
\begin{eqnarray}
  \Sigma^{R}_{\vec{Q},T}(\omega) & = & - \frac{\eta_{o} \hbar^{2} e^{2}}{m^{2}_{o}} \frac{|\chi_{ex}(0,\vec{Q})|^{2}}{E^{2}_{ex}(\vec{Q})}  \frac{\omega^{2}}{\sqrt{(Q c)^{2} - \omega^{2}}} \nonumber \\
  \Sigma^{R}_{\vec{Q},L}(\omega) & = &  \frac{\eta_{o} \hbar^{2} e^{2}}{m^{2}_{o}} \frac{|\chi_{ex}(0,\vec{Q})|^{2}}{E^{2}_{ex}(\vec{Q})} \sqrt{ (Q c)^{2} - \omega^{2}} \nonumber \\
\end{eqnarray}
The above expressions show that the longitudinal exciton will have a dispersion varying linearly with $Q$ only when $\hbar Qc >> E_{ex}(\vec{Q})$, and neither the longitudinal nor the transverse exciton will have a Dirac cone like dispersion for $\vec{Q} \approx 0$. After the completion of this work, we became aware that Gartstein et~al.~\cite{Gartstein15} have also reached a similar conclusion.

\subsection{Average Lifetime of a Thermal Exciton Ensemble: A Green's Function Approach} \label{app2}
The Green's function approach can also be used to find the average lifetime of a thermally distributed ensemble of excitons. The technique used below also incorporates, via the exciton spectral density function, exciton energy level broadening due to scattering processes. We define the exciton Green's function $G^{<}_{\vec{Q},s}(t-t')$ as follows~\cite{Mahan00},
\begin{equation}
G^{<}_{\vec{Q},s}(t-t') = \langle B^{\dagger}_{\vec{Q},s}(t') B_{\vec{Q},s}(t) \rangle
\end{equation}
The angular brackets indicate averaging with respect to a thermal ensemble of excitons. In the frequency domain,
\begin{equation}
 G^{<}_{\vec{Q},s}(\omega) = A_{\vec{Q},s}(\omega) n_{B}(\hbar\omega - \mu)   
\end{equation}
Here, $A_{\vec{Q},s}(\omega)$ is the spectral density function and $n_{B}(\hbar\omega - \mu)$ is the Boson occupation factor with chemical potential $\mu$. Most other exciton Green's functions can be obtained from the spectral density function~\cite{Mahan00}. The exciton density operator $\hat{n}_{ex}(t)$ is,
\begin{equation}
  \hat{n}_{ex}(t) = \frac{1}{A} \sum_{\vec{Q},s}  B^{\dagger}_{\vec{Q},s}(t) B_{\vec{Q},s}(t)
\end{equation}
The average density of excitons in thermal equilibrium can be written as,
\begin{equation}
  n_{ex} = \langle  \hat{n}_{ex}(t) \rangle = \frac{1}{A} \sum_{\vec{Q},s} \int \frac{d\omega}{2\pi} G^{<}_{\vec{Q},s}(\omega)
\end{equation}
To find the exciton radiative rate we use the density matrix perturbation theory assuming the exciton-radiation interaction Hamiltonian $\hat{H}_{int}$ as the perturbation. Since we are considering a resonant linear optical process, the interaction Hamiltonian given in (\ref{eq:light_H2}) or (\ref{eq:light_H3}) is adequate. We assume that the initial state of the excitons is described by a thermal density operator and that of the radiation by the vacuum state, and the initial joint density operator of the system is $\hat{\rho}_{i}$. In the standard interaction representation,
\begin{equation}
\frac{d\hat{n}^{I}_{ex}(t)}{dt} = -\frac{i}{\hbar} \left[ \hat{n}^{I}_{ex}(t),\hat{H}^{I}_{int}(t) \right]
\end{equation}
The result is,
\begin{equation}
  \left\langle  \frac{d\hat{n}_{ex}(t)}{dt} \right\rangle =  \frac{i}{\hbar} \int_{-\infty}^{t} dt' \, {\rm Tr}\left\{ \hat{\rho}_{i} \left[ \hat{H}^{I}_{int}(t') , \frac{d\hat{n}^{I}_{ex}(t)}{dt} \right] \right\} \nonumber \\
\end{equation}
The terms appearing inside the integral can be partitioned into appropriate radiation and exciton Green's functions. If one uses the following phenomenological form for the exciton spectral density function,
\begin{equation}
  A_{\vec{Q},s}(\omega) = \frac{2\hbar\Gamma_{ex}}{ (\hbar \omega - E_{ex}(\vec{Q}))^{2} + \Gamma^{2}_{ex}}
\end{equation}
then the final result can be written as,
\begin{eqnarray}
  & & \left\langle  \frac{d\hat{n}_{ex}(t)}{dt} \right\rangle = - \frac{1}{V} \sum_{\vec{q},s}  \left( \frac{e}{m_{o}} \right)^2 |\chi_{ex}(\vec{r}=0,\vec{q}_{\parallel})|^{2} \frac{1}{\epsilon_{o}\omega_{q}}  \nonumber \\
 & & \times \left(1 + \frac{q_{z}^{2}}{q^{2}} \right)  \frac{\Gamma_{ex}}{(E_{ex}(\vec{q}_{\parallel}) - \hbar \omega_{q})^{2} + \Gamma_{ex}^{2}}  n_{B}(\hbar\omega_{q} - \mu) \label{eq:ex_tau_3}
\end{eqnarray}
The ensemble average lifetime $\langle \tau_{sp} \rangle$ is then (see (\ref{eq:ensemble})),
\begin{equation}
  \frac{1}{\langle \tau_{sp} \rangle} = - \frac{1}{\langle \hat{n}_{ex}(t) \rangle}  \left\langle  \frac{d\hat{n}_{ex}(t)}{dt} \right\rangle \label{eq:tsp_app}
  \end{equation}
Note that in (\ref{eq:ex_tau_3}) the boson occupation factor is evaluated at the photon energy and not at the exciton energy. This subtlety could have been missed if the exciton radiative rate given earlier in (\ref{eq:ex_tau_2}) is averaged over the exciton density of states with a boson occupation factor evaluated at the exciton energy. Also note that the exciton density needs to be evaluated self-consistently using the same spectral density function. Since for Bosons the chemical potential must always be less than or equal to the lowest energy level, use of phenomenological Lorentzian spectral density functions can lead to problems during numerical calculations. A way around this problem is to use a spectral density function with tails that decay much faster than the tails of a Lorentzian and then the results for the radiative lifetimes do not depend sensitively on the exact form of the spectral density function. The result obtained using (\ref{eq:tsp_app}) is given in (\ref{eq:ensemble}). We find the same approximate result as that given in (\ref{eq:ensemble}) in both regimes, $\Gamma_{ex} << KT$ and $\Gamma_{ex} >> KT$, contrary to the findings of Citrin et~al.~\cite{Citrin93} in the case of III-V quantum well excitons. The difference in the results seems to originate from an incorrect evaluation of the exciton density in terms of the exciton spectral density function by Citrin et~al.~\cite{Citrin93}.


\begin{thebibliography}{99}

\bibitem{fai10} K. F. Mak, C. Lee, J. Hone, J. Shan, and T. F. Heinz, Phys. Rev. Lett. 105, 136805 (2010).
\bibitem{fai12} K. F. Mak, K. He, J. Shan, and T. F. Heinz, Nat. Nanotech. 7, 494 (2012).
\bibitem{fai13} K. F. Mak, K. He, C. Lee, G. H. Lee, J. Hone, T. F. Heinz, and J. Shan, Nat. Mater. 12, 207 (2013).
\bibitem{xu13}  J. S. Ross, S. Wu, H. Yu, N. J. Ghimire, A. M. Jones, G. Aivazian, J. Yan, D. G. Mandrus, D. Xiao, W. Yao, and X. Xu, Nat. Comm. 4, 1474 (2013). 	
\bibitem{wang10} A. Splendiani, L. Sun, Y. Zhang, T. Li, J. Kim, C. Chim, G. Galli, and F. Wang, Nano Lett. 10, 1271 (2010). 
\bibitem{kis11} B. Radisavljevic, A. Radenovic, J. Brivio, V. Giacometti, and A. Kis, Nat. Nanotech. 6, 147 (2011).
\bibitem{kis13} O. Lopez-Sanchez, D. Lembke, M. Kayci, A. Radenovic, and A. Kis, Nat. Nanotech. 8, 497 (2013).	
\bibitem{Changjian14} C. Zhang, H. Wang, W. Chan, C. Manolatou, F. Rana, Phys. Rev. B, 89, 205436 (2014). 
\bibitem{haugbook} H. Haug, S. W. Koch, {\em Quantum Theory of the Optical and Electronic Properties of Semiconductors}, World Scientific Publishing, Singapore (1990). 
\bibitem{timothy} T. C. Berkelbach, M. S. Hybertsen, and D. R. Reichman, Phys. Rev. B 88, 045318 (2013).
\bibitem{Chernikov14} A. Chernikov, T. C. Berkelbach, H. M. Hill, A. Rigosi, Y. Li, O. B. Aslan, D. R. Reichman, M. S. Hybertsen, T. F. Heinz, Phys. rev. Lett., 113, 076802 (2014). 
\bibitem{Lam12} T. Cheiwchanchamnangij and W. R. L. Lambrecht, Phys. Rev. B 85, 205302 (2012).
\bibitem{Louie1} D. Y. Qiu, F. H. da Jornada, and S. G. Louie, Phys. Rev. Lett. 111, 216805 (2013).
\bibitem{yao12} D. Xiao, Gui-Bin Liu, W. Feng, X. Xu, and W. Yao, Phys. Rev. Lett. 108, 196802 (2012).
\bibitem{Falko13} A. Kormanyos, V. Zolyomi, N. D. Drummond, P. Rakyta, G. Burkard and V. I. Falko, Phys. Rev. B 88, 045416 (2013). 
\bibitem{Andreani91} L. C. Andreani, F. Tassone, F. Bassani, Solid. Stat. Comm., 77, 641 (1991). 

\bibitem{Wang15b} H. Wang, J. H. Strait, C. Zhang, W. Chen, C. Manolatou, S. Tiwari, F. Rana, Phys. Rev. B 91, 165411 (2015). 
\bibitem{Siarkos00} A. Siarkos, E. Runge, and R. Zimmermann, Phys. Rev. B 61, 10854 (2000). 
  
\bibitem{Efimkin13} D. K. Efimkin, Yu. E. Lozovik, Phys. Rev. B, 87, 245416 (2013).  
\bibitem{Liu14} G. Liu, W. Shan, Y. Yao, W. Yao, D. Xiao, Phys. Rev., B, 88, 085433 (2014).  
\bibitem{Citrin93} D. S. Citrin, Phys. Rev. B, 47, 3832 (1993). 
\bibitem{Shi13} H. Shi, R. Yan, Rusen, S. Bertolazzi, J. Brivio, B. Gao, A. Kis, D. Jena, H. Xing, Huili, L. Huang, ACS Nano, 7, 1072 (2013).
\bibitem{Lagarde14} D. Lagarde, L. Bouet, X. Marie, C. R. Zhu, B. L. Liu, T. Amand, P H. Tan, B. Urbaszek, Phys. Rev. Lett., 112, 047401 (2014).
\bibitem{Korn11} T. Korn, S. Heydrich, M. Hirmer, J. Schmutzler, C. Schüller, App. Phys. Lett., 99, 102109 (2011).  

\bibitem{Tomimoto07} S. Tomimoto, A. Kurokawa, Y. Sakuma, T. Usuki, Y. Masumoto, Phys. Rev. B, 76, 205317 (2007).
\bibitem{Oberli97} D. Y. Oberli, F. Vouilloz, and E. Kapon, Phys. Stat. Sol., 164, 353 (1997).  
\bibitem{Kira06} M. Kira, S. W. Koch, Prog. Quant. Electron., 30, 155 (2006). 
\bibitem{Pepper00} D. Sanvitto, R. A. Hogg, A. J. Shields, D. M. Whittaker, M. Y. Simmons, D. A. Ritchie, M. Pepper, Phys. Rev. Lett., Phys. Rev. B, 62, R13294(R) (2000). 

\bibitem{Moody15} G. Moody, C. K. Dass, K. Hao, C.-H. Chen, L.-J. Li, A. Singh, K. Tran, G. Clark, X. Xu, G. Berghäuser, E. Malic, A. Knorr, X. Li, Nature Communications, 6, 8315 (2015).
\bibitem{Huber15} C. Poellmann, P. Steinleitner, U. Leierseder, P. Nagler, G. Plechinger, M. Porer, R. Bratschitsch, C. Schüller, T. Korn, R. Huber, Nature Materials, 14, 889 (2015).
\bibitem{Marie15} X. Marie, B. Urbaszek, Nature Materials, 14, 860 92015). 

\bibitem{Yao14} H. Yu, G. Liu, P. Gong, X. Xu, W. Yao, Nature Communications, 5, 3876 (2014).
\bibitem{Qiu15} D. Y. Qiu, T. Cao, S. G. Louie, Phys. Rev. Lett., 115, 176801 (2015).
  
\bibitem{Amani15} M. Amani, D. H. Lien, D. Kiriya, J. Xiao, A. Azcatl, J. Noh, S. R. Madhvapathy, R. Addou, S. KC, M. Dubey, K. Cho, R. M. Wallace, S. C. Lee, J. H. He, J. W. Ager III, X. Zhang, E. Yablonovitch, A. Javey, Science, 350 1065 (2015).
\bibitem{Mahan00} G. D. Mahan, {\em Many Particle Physics}, Springer, NY (2000).
\bibitem{Zhou15} J. Zhou, W. Y. Shan, W. Yao, Di Xiao, Phys. Rev. Lett., 115, 166803 (2015).
\bibitem{Srivastava15} A. Srivastava, A. Imamoglu, Phys. Rev. Lett., 115, 166802 (2015).

\bibitem{Girlanda95} S. Savasta, R. Girlanda, Solid State Communications, 96, 517 (1995). 
  
\bibitem{Guad80} E. Guadagnini, Il Nuovo Cimento, 57A, 294 (1980). 
\bibitem{PandA89} C. Cohen-Tannoudji, J. Dupont-Roc, G. Grynberg, {\em Photons and Atoms: Introduction to Quantum Electrodynamics}, Wiley, NY (1989).  

\bibitem{Gartstein15} Y. N. Gartstein, X. Li, C. Zhang, Phys. Rev. B, 92, 075445 (2015). 
  
\end{thebibliography}
\end{document}